\documentclass[11pt]{article}
\usepackage{epsfig}
\begin{document}

\title{\bf Critical state of phantom universe
\thanks{This work is supported by Liaoning
Province Educational Committee Research Project (20040026) and by
Bohai University.}}
\author{{\small Yi-Huan Wei}\\
{\small Department of Physics, Bohai University;} \\
{\small Institute of Physics, Bohai University, Jinzhou 121000,
Liaoning, China}\\} \maketitle

\begin{center}
{\bf Abstract}
\\ \vskip 0.5cm
\begin{minipage}{14cm}
\setlength{\baselineskip}{25pt plus1pt minus0pt} {\small ~~ The
late-time evolution behavior of the autonomous system in the
$SO(1,1)$ dark energy model with power-law potential is studied.
Big Rip may be a critical point of the autonomous system. This
means that such a Big Rip may be considered as the middle state
between the expanding and contracting phases of phantom universe.
This result is also valid for some special interactions between
matter and dark energy.}

\vskip 0.5cm
\end{minipage}
\end{center}
\vspace{0.5cm} \setlength{\baselineskip}{19pt plus0.8pt minus0pt}

\section {Introduction}

The observations \cite{Riess,Ber,AGRiess} indicates that the
universe is expanding acceleratedly and spatially flat, this
requires the existence of the dark energy
\cite{Rathra,Bento,MLi,Hu,Gong,Zhu}. Some dark energy models may
be related to the quintessence \cite{Boyle,Li,LHao}. The scalar
phantom model \cite{Cald} may be obtained from the quintessence
model by adopting a pure imaginary scale field. For the expanding
universe dominated by phantom ({\it phantom universe}), the energy
density increases with time. Phantom universe possesses some
appealing properties, which has greatly been studied
\cite{Cald,P,Chi,Elizalde,Abdalla,Caldwell,McInnes,Barboza,Dabrowski,Odintsov,Feng,Andrianov,Kujat}.
Some authors are interested in the dark energy models of two or
more scalar fields \cite{Wei,Guo,Zhang,Hao,Cai,WYH,Yihuan}, for
which the state of equation $w$ may change from $w>-1$ to $w<-1$.
We supposed the $SO(1,1)$ dark energy model \cite{Wei,WYH}, which
is relevant to the spintessence model having the $U(1)$ symmetry
\cite{Boyle}. Nevertheless, they are quite different in the
description for dark part in universe (dark energy or dark
matter). The former model may be considered as either a
quintessence-like or phantom model, while in the latter model the
state of equation is confined to be bigger than $-1$. We have
discussed the cases of the exponential and inverse power law
potentials, and find for both potentials there exist the late-time
phantom phases \cite{WYH,Yihuan}. For a phantom universe, there
can be the three different evolution consequences. First, if the
phantom equation of state tends rapidly to $-1$, the phantom
universe will approach the de Sitter universe. Secondly, it
encounters the future Big Rip. Thirdly, it has the future
singularity occurring at a finite scale factor studied in the
generalized phantom Chaplygin gas \cite{Bouhmadi}, in which the
energy density is given by $\rho=[A +
Ba^{3(1+\alpha)}]^{\frac{1}{1+\alpha}}$ with $B<0$.

We will focus on the phantom universe having a Big Rip. For it,
one can ask the following question. Is Big Rip the end of phantom
universe? Big Rip is a consequence of expanding phantom universe.
Generally, it is supposed to be a doomsday when all things will be
destroyed because the scale factor of universe becomes infinite
\cite{Caldwell}. In this paper, by analyzing the behavior of
phantom universe in the $SO(1,1)$ model we find that Big Rip may
correspond to the critical point of the autonomous system.
Considering that such a critical point is unstable, we suggest
that the Big Rip be the final state of an expanding phantom
universe and at the same time the initial state of the
corresponding contracting phantom universe.

\section {Critical point corresponding to Big Rip }

It was shown that in the $SO(1,1)$ model the phantom universe for
the inverse power law potential has a future Big Rip
\cite{Yihuan}. Here, we will further analyze the behavior of the
phantom universe near the Big Rip and infer that the Big Rip is a
critical point of the autonomous system.

Let us consider a spatially flat, isotropic and homogeneous
universe consisting of matter and dark energy. For this
background, the field equations read
\begin{eqnarray}
H^2=(\frac{\dot{a}}{a})^2=\frac{\kappa^2}{3}(\rho_{DE}+\rho_m),
\label{eq-1}
\end{eqnarray}
\begin{eqnarray}
\dot{H}=-\frac{\kappa^2}{2}(\rho_{DE}+\rho_m+p_{DE}+p_m),
\label{eq-2}
\end{eqnarray}
where $\rho_{DE}$ and $p_{DE}$ are energy and pressure density of
dark energy, $\rho_{m}$ and $p_{m}$ are those of matter,
$H=\frac{\dot{a}}{a}$ is the Hubble parameter, and a dot denotes
the derivative with respect to time. In the $SO(1,1)$ model
\cite{Wei}, the energy and pressure density of dark energy are
marked as $\rho_{\Phi}$ and $p_{\Phi}$, which are given by
\begin{eqnarray}
p_\Phi=\rho_k+\rho_c-V, \quad \rho_\Phi=\rho_k+\rho_c+V,
\label{eq-3}
\end{eqnarray}
where
\begin{eqnarray}
\rho_k=\frac{1}{2}\dot{\Phi}^2, \quad
\rho_c=-\frac{1}{2}\frac{\bar{Q}^2}{\Phi^2a^6}, \nonumber
\end{eqnarray}
with $\bar{Q}$ being the $SO(1,1)$ charge, and $V$ is a potential.
From Eq. (\ref{eq-3}), one can see that for $\rho_k<|\rho_c|$ dark
energy behaves as a phantom. Generically, for the matter part one
can consider the equation of state $p_m=(\gamma_m-1)\rho_m$ with
$0<\gamma_m\leq 2$ the barotropic index. For the universe
composing of dark energy and matter, the energy-momentum conserved
equation may be split into
\begin{eqnarray}
\dot{\rho}_\Phi+3H(\rho_\Phi+p_\Phi)=-C, \label{eq-4}
\end{eqnarray}
\begin{eqnarray}
\dot{\rho}_m+3H(\rho_m+p_m)=C, \label{eq-5}
\end{eqnarray}
where $C$ denotes the interaction between matter and dark energy.

Introducing the dimensionless variables
\begin{eqnarray}
x\equiv\pm\frac{\kappa \sqrt{\rho_k}}{\sqrt{3}H}=\frac{\kappa
\dot{\Phi}}{\sqrt{6}H}, \nonumber \\
y\equiv\frac{\kappa \sqrt{V}}{\sqrt{3}H}, \quad
z\equiv\frac{\kappa
\sqrt{\rho_m}}{\sqrt{3}H}, \label{eq-6} \\
w\equiv\frac{\kappa \Phi}{\sqrt{6}}, \quad v\equiv\frac{\kappa
\sqrt{-\rho_c}}{\sqrt{3}H}, \nonumber
\end{eqnarray}
then Eqs. (\ref{eq-1})-(\ref{eq-5}) can be rewritten as an
autonomous system \cite{Cai}
\begin{eqnarray}
x'=3x(x^2-v^2+\frac{\gamma_m}{2}z^2-1)-w^{-1}v^2-\frac{\kappa
V_{,\phi}}{\sqrt{6}H^2}-C_1, \quad \nonumber
\end{eqnarray}
\begin{eqnarray}
y'=3y(x^2-v^2+\frac{\gamma_m}{2}z^2)+\frac{\kappa}{2\sqrt{3}H}\frac{V_{,\phi}}{\sqrt{V}}\frac{\dot{\phi}}{H},
\quad\nonumber
\end{eqnarray}
\begin{eqnarray}
z'=3z(x^2-v^2+\frac{\gamma_m}{2}z^2-\frac{\gamma_m}{2})+C_2,
\quad\label{eq-system}
\end{eqnarray}
\begin{eqnarray}
w'=x, \quad\nonumber
\end{eqnarray}
\begin{eqnarray}
v'=3v(x^2-v^2+\frac{\gamma_m}{2}z^2-1)-xw^{-1}v, \quad\nonumber
\end{eqnarray}
where $C_1=\frac{\kappa C}{\sqrt{6}H^2\dot{\Phi}}$ and
$C_2=\frac{\kappa C}{2\sqrt{3\rho_m}H^2}$, a prime denotes the
derivative with respect to $\mathcal{N}=\ln a$ and a dot the
derivative with respect to $\phi$.

On the basis of work in Ref. \cite{Yihuan}, we will proceed to
discuss the $SO(1,1)$ model with the inverse power law potential
\begin{eqnarray}
V=V_0\Phi^{-n}, \label{eq-8}
\end{eqnarray}
where $0<n<2$ and $V_0$ are two constants. For this potential, the
autonomous system (\ref{eq-system}) reduces to
\begin{eqnarray}
x'=3x(x^2-v^2+\frac{\gamma_m}{2}z^2-1)+w^{-1}(\frac{n}{2}y^2-v^2)-C_1,
\quad\label{eq-9}
\end{eqnarray}
\begin{eqnarray}
y'=3y(x^2-v^2+\frac{\gamma_m}{2}z^2)-\frac{n}{2}w^{-1}xy,
\quad\label{eq-10}
\end{eqnarray}
\begin{eqnarray}
z'=3z(x^2-v^2+\frac{\gamma_m}{2}z^2-\frac{\gamma_m}{2})+C_2,
\quad\label{eq-11}
\end{eqnarray}
\begin{eqnarray}
w'=x, \quad\label{eq-12}
\end{eqnarray}
\begin{eqnarray}
v'=3v(x^2-v^2+\frac{\gamma_m}{2}z^2-1)-xw^{-1}v.
\quad\label{eq-13}
\end{eqnarray}

For the autonomous system (\ref{eq-9})-(\ref{eq-13}), we find the
following critical point
\begin{eqnarray}
(\bar{x},\bar{y},\bar{z},\bar{w},\bar{v})=(0,\sqrt{\frac{2}{2-n}},0,0,\sqrt{\frac{n}{2-n}}),
\quad\label{eq-14}
\end{eqnarray}
with $\bar{w}=0$ and $\bar{x}=0$ satisfying
\begin{eqnarray}
\bar{w}^{-1}\bar{x}=-\frac{6}{n} \bar{v}^2,  \quad\label{eq-15}
\end{eqnarray}
where a bar denotes the values of the dimensionless variables at
critical point.

In order to check the critical point (\ref{eq-14}), we start from
Eq. (\ref{eq-4}) which contains a interaction term $C$. For the
inverse power law potential, (\ref{eq-4}) reduces to
\begin{eqnarray}
\ddot{\Phi}+3H\dot{\Phi}+\bar{Q}^2\Phi^{-3}a^{-6}-nV_0\Phi^{-(n+1)}=-C\dot{\Phi}^{-1},
\label{conserv}
\end{eqnarray}
or
\begin{eqnarray}
a=[\frac{\bar{Q}^2}{nV_0}(1+\eta_1)\Phi^{n-2}]^{\frac{1}{6}},
\label{scale}
\end{eqnarray}
with
\begin{eqnarray}
\eta_1=\bar{Q}^{-2}(\ddot{\Phi}+3H\dot{\Phi}+C\dot{\Phi}^{-1})a^{6}\Phi^{3}.
\label{eta}
\end{eqnarray}
According to Eq. (\ref{scale}), the Hubble parameter is written as
\begin{eqnarray}
H=\frac{n-2}{6}\frac{\dot{\Phi}}{\Phi}+\frac{\dot{\eta}_1}{6(1+\eta_1)}.
\label{hubble}
\end{eqnarray}
Combining Eq. (\ref{eq-1}) and (\ref{hubble}) gives rise to
\begin{eqnarray}
-\frac{\dot{\Phi}}{\Phi}+\xi=\frac{2\sqrt{3}}{2-n}\delta \kappa
V^{\frac{1}{2}}\sqrt{1+\frac{\eta_3}{\delta^{2}}}, \label{eq-phi}
\end{eqnarray}
where
\begin{eqnarray}
\delta=[1-\frac{n(1-\eta_2)}{2(1+\eta_1)}]^{\frac{1}{2}},
\label{parameter1}
\end{eqnarray}
\begin{eqnarray}
\quad \xi=\frac{\dot{\eta}_1}{(2-n)(1+\eta_1)}, \label{parameter2}
\end{eqnarray}
\begin{eqnarray}
\eta_2=-\frac{\rho_k}{\rho_c}, \quad \eta_3=\frac{\rho_m}{V}.
\label{parameter3}
\end{eqnarray}

Let us first consider the $C=0$ case. As has been done in Refs.
\cite{WYH,Yihuan}, we assume the parameters $\eta_1$, $\eta_2$ and
$\eta_3$ to be some small quantities at late times. Under this
assumption, the late-time field and scale factor are the same as
those given in \cite{Yihuan}
\begin{eqnarray}
\Phi_L\simeq\Phi_0\tau^{\frac{2}{n}}, \quad
\Phi_0=[\frac{3n^2k^2V_0}{2(2-n)}]^{\frac{1}{n}}, \label{LP}
\end{eqnarray}
\begin{eqnarray}
a_L\simeq A_0[\bar{Q}\tau^{\frac{n-2}{n}}]^{\frac{1}{3}}, \quad
A_0=(\frac{1}{nV_0}\Phi_0^{n-2})^{\frac{1}{6}}, \label{LA}
\end{eqnarray}
with $\tau=t_{br}-t$, where a subscript "{\it{L}}" denotes the
meaning of {\it late-time}. From (\ref{LP}) and (\ref{LA}), follow
the late-time potential $V$ and $\rho_{c}$
\begin{eqnarray} V_L\simeq V_0\Phi^{-n}\simeq
\frac{2(2-n)}{3n^2\kappa^{2}}\tau^{-2}, \label{LV}
\end{eqnarray}
\begin{eqnarray}
\rho_{cL}\simeq-\frac{\bar{Q}^2}{2\Phi^2 a^6}\simeq
\frac{n-2}{3n\kappa^{2}}\tau^{-2}. \label{LC}
\end{eqnarray}
According to $\rho_m\sim a^{-3\gamma_m}$, then we have the
late-time matter density
\begin{eqnarray}
\rho_{mL}=\rho_{m0}\bar{Q}^{-\gamma_m}\tau^{\frac{2-n}{n}\gamma_m},
\label{LM}
\end{eqnarray}
where $\rho_{m0}$ is a constant.

From Eq. (\ref{eq-6}) and Eqs. (\ref{LP})-(\ref{LM}), we obtain
the late-time $x$, $y$, $z$, $w$ and $v$ as
\begin{eqnarray}
x_L\simeq
-\frac{\sqrt{6}k}{2-n}[\frac{3n^2k^2V_0}{2(2-n)}]^{\frac{1}{n}}\tau^{\frac{2}{n}},
\label{eq-Lx}
\\
y_L\simeq \sqrt{\frac{2}{2-n}}, \label{eq-Ly} \\
\quad z_L\simeq
\frac{\sqrt{3}n}{2-n}kB_0^{\frac{1}{2}}\bar{Q}^{-\frac{\gamma_m}{2}}
\tau^{\frac{2-n}{2n}\gamma_m+1}, \label{eq-Lz} \\
w_L\simeq\frac{k}{\sqrt{6}}[\frac{3n^2k^2V_0}{2(2-n)}]^{\frac{1}{n}}\tau^{\frac{2}{n}}, \label{eq-Lw} \\
v_L\simeq \sqrt{\frac{n}{2-n}}. \label{eq-Lv}
\end{eqnarray}
Eqs. (\ref{eq-Lx})-(\ref{eq-Lv}) tell the fact that at $\tau=0$
there is
$(\bar{x},\bar{y},\bar{z},\bar{w},\bar{v})=(0,\sqrt{\frac{2}{2-n}},0,0,\sqrt{\frac{n}{2-n}})$.
In what follows, we confirm it a critical point.

Noting that $x'=\dot{x}H^{-1}$ and the late-time Hubble parameter
$H_L\simeq \frac{2-n}{3n}\tau^{-1}$, from Eqs.
(\ref{eq-9})-(\ref{eq-13}) we have
\begin{eqnarray}
x'_L\simeq
-\frac{6\sqrt{6}k}{(n-2)^2}[\frac{3n^2k^2V_0}{2(2-n)}]^{\frac{1}{n}}\tau^{\frac{2}{n}}\simeq
0, \label{eq-Px}
\\
y'_L\simeq 0, \label{eq-Py} \\
z'_L\simeq C
\tau^{\frac{2-n}{2n}\gamma_m+1}\simeq 0, \label{eq-Pz} \\
w'_L\simeq\frac{\sqrt{6}k}{2-n}[\frac{3n^2k^2V_0}{2(2-n)}]^{\frac{1}{n}}\tau^{\frac{2}{n}}\simeq 0, \label{eq-Pw} \\
v'_L\simeq 0. \label{eq-Pv}
\end{eqnarray}
One may also start from Eqs. (\ref{eq-Lx})-(\ref{eq-Lv}) with
$C_1=0$ and $C_2=0$ to derive Eqs. (\ref{eq-Px})-(\ref{eq-Pv}).
From Eqs. (\ref{eq-Px})-(\ref{eq-Pv}), it is deduced that at
$\tau=0$ there exactly is
$(\bar{x}',\bar{y}',\bar{z}',\bar{w}',\bar{v}')=(0,0,0,0,0)$. A
Big Rip corresponding to the critical point of the autonomous
system can have an important hint for the evolution of phantom
universe. A discussion for this will be given in the end of the
next section.

\section {Discussions}

In this section, we first show that for the two cases of
interactions the Big Rip will also occur, which corresponds to the
critical point of the autonomous system. Then, we give a possible
discussion of the phantom universe with Big Rip corresponding to
the critical point.

Let us assume that (\ref{LP}) and (\ref{LA}) are still appropriate
for some special interactions. Under this assumption, we may
proceed to determine the three parameters $\eta_1$, $\eta_2$ and
$\eta_3$. From (\ref{eta}) and (\ref{parameter3}) we get

\begin{eqnarray}
\eta_1\simeq C\bar{Q}^{-2}\dot{\Phi}_L^{-1}a_L^{6}\Phi_L^{3}\sim
C\tau^3, \label{eq-E1}
\end{eqnarray}
\begin{eqnarray}
\eta_2\simeq\frac{6}{n(2-n)}\Phi_0^2k^2\tau^{\frac{4}{n}},
\label{eq-E2}
\end{eqnarray}
\begin{eqnarray}
\eta_3=\frac{\rho_{mL}}{V}\simeq \frac{3n^2k^2}{2(2-n)}B_0
\bar{Q}^{-\gamma_m}\tau^{\frac{2-n}{n}\gamma_m+2}. \label{eq-E3}
\end{eqnarray}
Clearly, $\eta_2$ and $\eta_3$ tend to zero with $\tau\rightarrow
0$, noting $0<n<2$ and $\gamma_m>0$. The late-time behavior of
$\eta_1$ depends on that of $C$. Providing that $C$ behaves as
$C\sim\tau^p$ with $p>-3$ at late times, then $\eta_1$ is a small
quantity. One can ask the question that such $\eta_1$ doesn't
guarantee the quantity $\xi$ in (\ref{eq-phi}) is always a small
quantity. It is indeed the case for $p\leq-2$. Nevertheless, it is
enough for obtaining (\ref{LP}) and (\ref{LA}) only if $\xi$ is a
relative small quantity to the term $\dot{\Phi}/\Phi\sim
\tau^{-1}$.

Next, we find the interactions that satisfy the requirement,
$p>-3$. Some forms of interactions have been derived from string
theory and scalar-tensor theory \cite{Luca,Damour,CHao}. Here, we
will analyze the following two cases: (I)
$C=\alpha\kappa\rho_m\dot{\Phi}$ and (II) $C=3\beta H\rho_m$ with
$\alpha$ and $\beta$ two positive constants. Noting that
$\dot{\Phi}_L\sim\tau^{\frac{2}{n}-1}$ and
$\rho_m\sim\tau^{\frac{2-n}{n}\gamma_m}$ with $0<n<2$ and
$\gamma_m>0$, one can see the interaction (I) satisfies the
requirement for $p$ obviously.

Compared to (I), the interaction (II) is stronger and the analysis
for it is slightly complicated. For this interaction, Eqs.
(\ref{eq-4}) and (\ref{eq-5}) may be rearranged as
\begin{eqnarray}
\dot{\rho}_\Phi+3H\tilde{\gamma}_\Phi\rho_\Phi=0, \quad
\tilde{\gamma}_\Phi=\gamma_\Phi+\beta r \label{InterD}
\end{eqnarray}
with $r=\frac{\rho_m}{\rho_\Phi}$ and the late-time
$\gamma_\Phi=1+w_\Phi\simeq-\frac{2n}{2-n}$, and
\begin{eqnarray}
\dot{\rho}_m+3H\tilde{\gamma}_m\rho_m=0, \quad
\tilde{\gamma}_m=\gamma_m-\beta, \label{InterM}
\end{eqnarray}
which is the same as the one satisfied by the noninteracting
matter, with only $\gamma_m$ being replaced by $\tilde{\gamma}_m$.
For the parameter $r$, it is a small quantity at late times, only
if $\tilde{\gamma}_m>\tilde{\gamma}_\Phi$. For $z$ being a small
quantity at late times, it is required that
$\tilde{\gamma}_m>-\frac{2n}{2-n}$. If one further requires matter
density to decrease with time, i.e., $\tilde{\gamma}_m>0$, then
the lower limit of $\gamma_m$ should be modified to $\beta$ with
$\beta<2$. In the following we will consider this case.

In the case of interaction (II), we need to reanalyze Eqs.
(\ref{eq-9}) and (\ref{eq-11}) to determine whether the Big Rip is
proceed to be a critical point. Defining
$\tilde{C}=w^{-1}(\frac{n}{2}y^2-v^2)$, then directly starting
from the definitions of $w$, $y$ and $v$ gives rise to
\begin{eqnarray}
\tilde{C}_L=\frac{\kappa
nV_0}{\sqrt{6}H^2}\Phi_L^{-(n+1)}\eta_1(1+\eta_1)^{-1}.
\label{eq-44}
\end{eqnarray}
Substituting $\eta_1\simeq
C\bar{Q}^{-2}\dot{\Phi}_L^{-1}a_L^{6}\Phi_L^{3}\simeq\frac{1}{nV_0}\dot{\Phi}_L^{-1}\Phi_L^{n+1}C$
into the numerator of the factor $\frac{\eta_1}{1+\eta_1}$ on the
right hand-side of Eq. (\ref{eq-44}) yields $\tilde{C}_L\simeq
\frac{1}{1+\eta_1} C_{1L}$ with $C_{1L}=\frac{\kappa
C}{\sqrt{6}H_L^2\dot{\Phi}_L}$, which gives
\begin{eqnarray}
\tilde{C}_L-C_{1L}\simeq\frac{\kappa}{\sqrt{6}nV_0}H_L^{-2}\dot{\Phi}_L^{-2}\Phi_L^{n+1}C^2.
\label{eq-45}
\end{eqnarray}
Noting that $C\sim\tau^{-1+\frac{2-n}{n}\tilde{\gamma}_m}$ at late
times, then there is
$\tilde{C}_L-C_{1L}\sim\tau^{4-\frac{2}{n}+\frac{2(2-n)}{n}\tilde{\gamma}_m}$
near Big Rip time. Setting $\tilde{C}_L-C_{1L}=0$ at $\tau=0$
leads to the following inequality
\begin{eqnarray}
n(2-\tilde{\gamma}_m)-1+2\tilde{\gamma}_m>0. \label{eq-46}
\end{eqnarray}
For $\tilde{\gamma}_m\geq\frac{1}{2}$ inequality (\ref{eq-46}) can
always be realizable, while for $\tilde{\gamma}_m<\frac{1}{2}$ the
constraint on $n$ should be imposed,
$n>\frac{1-2\tilde{\gamma}_m}{2-\tilde{\gamma}_m}$. Noting that
$0<n<2$ and $0<\tilde{\gamma}_m\leq2$, then one can easily find
$C_{2L}=0$ at $\tau=0$. As a result, for the interaction (II) with
$n$ and $\tilde{\gamma}_m$ satisfying (\ref{eq-46}), the Big Rip
can still be the critical point. When (\ref{eq-46}) isn't
satisfied, the Big Rip will be no longer a critical point. This
means there are the two different kinds of Big Rip, which may be
distinguished from whether they are a critical point. Here, we are
interested in the case that Big Rip corresponding to the critical
point and attempt to give a possible interpretation for it.

The critical point discussed here is obviously unstable since Big
Rip is only an evolution consequence of expanding phantom
universe. What does a Big Rip mean? Generically, it is assumed to
be the final state of an expanding phantom universe. In Ref.
\cite{Yihuan}, we have demonstrated the symmetric properties of
the field and scale factor of phantom universe about the Big Rip.
This suggests that the solution of the phantom universe having a
Big Rip should be extended to the range $\tau<0$. In this paper,
we have shown that the Big Rip in the $SO(1,1)$ model can be a
critical point. For such phantom universe that the Big Rip is a
critical point we conjecture the Big Rip may only be a middle
point in its evolving. In other words, such a phantom universe
should undergo the two evolution phases, the accelerating
expansion and decelerating contraction which transit at the Big
Rip.

\end{document}